\documentstyle[12pt,epsfig]{article}
\def\be{\begin{equation}}
\def\ee{\end{equation}}
\def\bear{\be\begin{array}}
\def\eear{\end{array}\ee}
\def\bea{\begin{eqnarray}}
\def\eea{\end{eqnarray}}

\def\baselinestretch{1}
\begin{document}
\catcode`@=11
\newtoks\@stequation
\def\subequations{\refstepcounter{equation}%
\edef\@savedequation{\the\c@equation}%
  \@stequation=\expandafter{\theequation}
  \edef\@savedtheequation{\the\@stequation}
  \edef\oldtheequation{\theequation}%
  \setcounter{equation}{0}%
  \def\theequation{\oldtheequation\alph{equation}}}
\def\endsubequations{\setcounter{equation}{\@savedequation}%
  \@stequation=\expandafter{\@savedtheequation}%
  \edef\theequation{\the\@stequation}\global\@ignoretrue

\noindent}
\catcode`@=12
\begin{titlepage}
\title{{\bf An analysis of soft terms in Calabi-Yau 
compactifications}\thanks{Research supported in part by: the CICYT, under
contract AEN93-0673 (CM); the European Union,
under contracts CHRX-CT93-0132 (CM) and
SC1-CT92-0792 (CM); the
Ministerio de Educaci\'on y Ciencia, under research grant (HBK).}
}
\author{{\bf H.B. Kim}
and {\bf C. Mu\~noz}\\
\hspace{3cm}\\
{\small Departamento de F\'{\i}sica
Te\'orica C--XI} \\
{\small Universidad Aut\'onoma de Madrid, 28049 Madrid, Spain}\\
{\small hbkim@delta.ft.uam.es $$ cmunoz@ccuam3.sdi.uam.es}}
\date{}
\maketitle
\def\baselinestretch{1.15}
\begin{abstract}
\noindent

We perform an analysis of the soft supersymmetry-breaking terms arising
in Calabi-Yau compactifications. The sigma-model contribution and the
instanton correction to the K\"ahler potential are included in the 
computation. The existence of off-diagonal moduli and matter metrics gives
rise to specific features as the possibility of 
having scalars heavier than gauginos or the presence of tachyons. 
Although non-universal soft terms is a natural situation, we point out that 
there is an interesting limit where universality is achieved.
Finally, we
compare these results with those of orbifold compactifications. 
Although they are qualitatively similar some features indeed change. For
example, sum rules found in orbifold models which imply that on average
the scalars are lighter than gauginos can be violated in Calabi-Yau
manifolds.

\end{abstract}

\thispagestyle{empty}

\leftline{}
\leftline{}
\leftline{FTUAM 96/29}
\leftline{hep-ph/9608214}
\leftline{July 1996}

\vskip-20.cm
\rightline{ FTUAM 96/29}
\rightline{hep-ph/9608214}
\vskip3in

\end{titlepage}
\newpage
\setcounter{page}{1}

\section{Introduction}

The computation of the soft Supersymmetry (SUSY)-breaking terms in
effective N=1 theories coming from four-dimensional (4-D) Superstrings 
is crucial in order to connect these theories with the low-energy 
phenomena \cite{Munoz1}. The soft terms not only 
contribute to the Higgs
potential generating the radiative breakdown of the electroweak symmetry,
but also
determine the SUSY spectrum, like gaugino, squark or slepton masses.

The most extensive studies of soft terms, working
at the perturbative 
level\footnote{For possible non-perturbative string effects
on soft terms see ref.\cite{Alberto}.}, 
have been
carried out in the context of (0,2) symmetric Abelian 
orbifolds \cite{orbifolds}
assuming that the seed of SUSY breaking is located in the
dilaton/moduli sectors [4-9]\footnote{The limit where only the dilaton
contributes to SUSY breaking is compactification-scheme independent and was
analyzed in \cite{Kaplunovsky,Brignole,Ross,Alberto}. 
Its phenomenology was studied
in refs.\cite{Barbieri,Brignole}.}.
The latter implies that the Goldstino field is a linear combination of the
fermionic partners of the dilaton $S$, the field whose vacuum expectation
value (VEV) determines the gauge coupling constant, and the moduli
$T_i$, $U_j$, the fields whose VEVs parametrize the size and shape of
the compactified space.
In particular, in refs.\cite{Brignole,Brignole2}, 
where no special assumption was made
about the possible origin of SUSY breaking, the soft terms depend on
the gravitino mass $m_{3/2}$ and the parameters specifying the Goldstino
direction. This simplify the analysis of soft terms and leads to some
interesting relationships among themselves which could perhaps be
experimentally 
tested \cite{Brignole,japoneses,Brignole2,Savoy}.

It would be interesting to extend this approach for more complicated
compactifications as Calabi-Yau (CY) manifolds \cite{CY}.
This is in fact the aim of the present paper\footnote{Some
computations of soft terms in CY compactifications were also carried out 
in refs.\cite{Brignole,Ferrara}. In particular, in \cite{Brignole} only the
``overall modulus'' case was studied whereas in \cite{Ferrara} non-perturbative
effects breaking SUSY were considered in order to cancel the tree-level and 
one-loop cosmological constant.}. Unfortunately the
state-of-the-art CY technology will limit our computation of soft terms.
First, the analysis of (0,2) vacua is an open problem in CY compactifications
and therefore we are forced to restrict our study to (2,2) theories. In
this type of theories the gauge group is $E_6\times E_8$ and several families
of ${\bf 27}$ and $\overline{\bf 27}$ 
matter fields can be present. May be these are not
the most realistic low-energy theories since usually one likes gauge groups
smaller than $E_6$. 
In any case we hope, as occurs in orbifolds, 
they give us an insight into features of (0,2) models.
On the other hand, if the suggestion that all SUSY (2,2)
vacua with $E_6\times E_8$ gauge symmetry are compactifications on CY
manifolds \cite{Cual} is correct, our results would be of interest for any
compactification scheme with those properties.
Finally, we will focus on the region of the moduli space where all the
${\rm Re} T_i$ are large, often called the large-radius limit.
The reason is that this limit provides a good approximation to the 4-D
effective theory. The extension to small manifolds is quite involved and again
an open problem. Large $T$, in practice, does not really mean 
$T\rightarrow \infty$, since the world-sheet instanton corrections are
exponentially suppressed. For values $|T|\geq 2-3$ these world-sheet instanton
contributions can often be neglected and, in this sense these $T$-values
are already large. On the other hand, the VEV of the moduli are not
necessarily of order one. As studied in specific orbifold examples in
ref.\cite{unificacion}, 
${\rm Re} T\simeq 5-10$ is mandatory if the discrepancy between
the unification scale of the gauge couplings and the string
unification scale is explained by the effect of string threshold corrections.
This might also be the case of CY compactifications.

The structure of the paper is as follows. In section 2 we discuss 
some formulae for the computation of 
soft terms in effective Supergravity (SUGRA) theories from Strings, in the
most general case of off-diagonal metrics. In the spirit of
refs.\cite{Brignole,Brignole2}, i.e. just assuming that SUSY is broken by
dilaton/moduli fields, we parametrize the SUSY breaking in terms
of ``Goldstino angles'' 
generalizing the ones used there
to the case of off-diagonal
moduli metric. This is precisely the case of CY compactifications.
Unlike orbifolds, 
where off-diagonal metrics are relatively rare, these are
present in general in CY manifolds.
Thus in section 3 we apply
the formulae obtained in section 2
to the computation of the soft terms.
After summarizing some generic features of effective SUGRAs
coming from CY compactifications, we 
obtain general formulae for the soft terms which are independent of the
size of the manifold. 
An interesting limit where universality is achieved is studied in 3.2. Then, 
in order to obtain more concrete features we work
in the large-radius limit, where the prepotential is known, computing
the K\"ahler potential
including the sigma-model and instanton contributions. First we 
discuss the 
simple cases of CY manifolds with only one
K\"ahler modulus. These will be good guiding examples in order 
to study in detail
the modifications produced on soft terms by the above-mentioned 
contributions\footnote{The importance of considering this type of
contributions was first pointed out in ref.\cite{Kiwoon}.}.
Second, 
we consider the
general case of CY compactifications with several 
moduli studying what new features appear. Finally, in section 4,
we compare the results on soft terms obtained in the previous sections with
those of orbifold compactifications. We leave the conclusions for section 5.

\section{General parametrization of soft terms}

We are going to consider $N=1$ SUSY 4-D Strings with
$m$ moduli $T_i$, $i=1,..,m$. Such notation refers to both $T$-type
and $U$-type (K\"ahler class and complex structure in the Calabi-Yau
language) fields.
In addition there will be charged matter fields $C_{\alpha }$ and the
dilaton field $S$.
The associated effective $N=1$ SUGRA
K\"ahler potentials (to first order in the matter fields) are of the
type:
\begin{eqnarray}
K(S,S^*,T_i,T_i^*,C_{\alpha},C_{\alpha}^*) &=&
-\log(S+S^*)\ +\ {\hat K}(T_i,T_i^*)
+\ {\tilde K}_{{\overline{\alpha}}{\beta}}(T_i,T_i^*)
{C^*}^{\overline{\alpha}} C^{\beta }\ 
\nonumber\\ &&
+\ (Z_{{\alpha }{ \beta }}(T_i,T_i^*){C}^{\alpha}
C^{\beta }\ +\ h.c. \ )\ .
\label{kahl}
\end{eqnarray}
The first piece is the usual term corresponding to the 
dilaton $S$ which is present for any compactification whereas the
second is the K\"ahler potential of the moduli fields. 
The greek indices label the matter fields and their
kinetic term functions are  given by
${\tilde K_{{\overline{\alpha }}{ \beta }}}$ and $Z_{{\alpha }{\beta }}$
to lowest order in the matter fields. The last piece is often forbidden
by gauge invariance in specific models although it may be relevant
in order to solve the $\mu$ problem \cite{GM,Kaplunovsky}.
The complete $N=1$ SUGRA Lagrangian is determined by
the K\"ahler potential $K({\phi }_M ,\phi^*_M)$, the superpotential
$W({\phi }_M)$ and the gauge kinetic functions
$f_a({\phi }_M)$, where $\phi_M$ generically denotes the chiral fields
$S,T_i,C_{\alpha }$. As is well known, $K$ and $W$ appear in the
Lagrangian only in the combination $G=K+\log|W|^2$. In particular,
the ($F$ part of the) scalar potential is given by
\begin{eqnarray}
& V(\phi _M, \phi ^*_M)\ =\
e^{G} \left( G_M{K}^{M{\bar N}} G_{\bar N}\ -\ 3\right) =  
\left( F^{\bar N}{K}_{{\bar N}M} F^M\ -\ 3e^{G}\right)
\ , &
\label{pot}
\end{eqnarray}
where $G_M \equiv \partial_M G \equiv \partial G/ \partial \phi_M$ 
and $K^{M{\bar N}}$ is the inverse of the K\"ahler metric
$K_{{\bar N }M}\equiv{\partial}_{\bar N}{\partial }_M K$.
We have also written $V$ as a function of the $\phi_M$  auxiliary fields, 
$F^M=e^{G/2} {K}^{M {\bar P}} G_{\bar P}$.

Following the spirit of refs.\cite{Brignole,Brignole2}
the crucial assumption  now is  to locate the origin of SUSY breaking in the
dilaton/moduli sector. 
Then, applying the
standard SUGRA formulae \cite{Soni,Kaplunovsky} 
to the most general case when the
moduli and matter metrics are not diagonal we obtain:
\begin{eqnarray}
{m}'^2_{{\overline{\alpha }}{ \beta }} &=& 
m_{3/2}^2 {\tilde K_{{\overline{\alpha }}{ \beta }}}
- {F}^{\overline{i}} ( \partial_{\overline{i}}\partial_j
{\tilde K_{{\overline{\alpha }}{ \beta }}}
-\partial_{\overline{i}} {\tilde K_{{\overline{\alpha }}{ \gamma}}}
{\tilde K^{{ \gamma} {\overline{\delta}} }}
\partial_j {\tilde K_{{\overline{\delta}}{ \beta}}}  ) F^j \ , 
\label{mmatrix}
\\
A'_{\alpha\beta\gamma} &=& 
F^S K_S Y'_{\alpha\beta\gamma}
\nonumber\\ &&
+ F^i \left[\frac12{\hat K}_i Y'_{\alpha\beta\gamma}
+ \partial_i Y'_{\alpha\beta\gamma} - \left\{
{\tilde K^{{ \delta} {\overline{\rho}} }}
\partial_i {\tilde K_{{\overline{\rho}}{ \alpha}}} Y'_{\delta\beta\gamma}
+(\alpha \leftrightarrow \beta)+(\alpha \leftrightarrow \gamma)\right\}
\right]\ ,
\label{mmatrix2}
\end{eqnarray}
where ${m}'^2_{ {\overline{\alpha }} { \beta } }$
and $A'_{\alpha\beta\gamma}$ are the soft mass matrix and the soft 
trilinear parameters respectively (corresponding to un-normalized charged
fields), $m_{3/2}^2=e^G$ is the gravitino mass-squared, 
$Y'_{\alpha \beta \gamma }\equiv e^{K/2}Y_{\alpha \beta \gamma}$ with 
$Y_{\alpha \beta \gamma}$ a 
renormalizable
Yukawa coupling involving three charged chiral fields 
and finally, 
$F^S=e^{G/2} K_{ {\bar{S}} S}^{-1} G_{\bar{S}}$, 
$F^i=e^{G/2} {\hat K}^{i {\overline j}} G_{\overline j}$ are the dilaton and
moduli auxiliary fields respectively. 
Notice that, as discussed in \cite{Brignole2}, 
after normalizing the fields to get
canonical kinetic terms, the first piece in
eq.(\ref{mmatrix}) will lead to universal diagonal soft
masses but the second piece will generically induce
off-diagonal contributions. Concerning the
$A$-parameters, notice that we have not
factored out the Yukawa couplings as usual, since
proportionality is not guaranteed.
Indeed, although the first term in
$A'_{\alpha\beta\gamma}$ is always proportional
in flavour space to the corresponding Yukawa
coupling, the same thing is not necessarily true
for the other terms.
We will see below that this is precisely what occurs in general with 
scalar masses and trilinear parameters in
CY compactifications. However, there is an interesting limit where
universality is achieved.

If the VEV of the scalar potential eq.(\ref{pot}),
$V_0$, is not assumed to be zero one just has to replace
$m_{3/2}^2\rightarrow m_{3/2}^2+V_0$   
in the expression of ${m}'^2_{ {\overline{\alpha }} { \beta } }$.

Physical gaugino masses $M_a$ for the canonically normalized gaugino fields
are given by 
\begin{equation}
M_a=\frac12\left({\rm Re\:}f_a\right)^{-1}
\left[F^S\partial_Sf_a+F^i\partial_i f_a\right]\ .
\label{gauginos}
\end{equation}

Let us take the following parametrization
for the VEVs of the dilaton and moduli auxiliary
fields
\begin{eqnarray}
F^S &=& \sqrt3m_{3/2}\sin\theta (S+S^*)e^{-i\gamma_S}\ , \nonumber\\
F^i &=& \sqrt3m_{3/2}\cos\theta P^{i{\bar j}}\Theta_{\bar j}\ ,
\label{parametrization}
\end{eqnarray}
which generalize the one used in ref.\cite{Brignole2} to the
case of off-diagonal moduli metric\footnote{We 
note for completeness that in ref.\cite{Brignole2} some orbifold models
with off-diagonal metrics were also studied. The
moduli and matter metrics were
${\hat K}_{ {\overline i} j}=t^{-1}_{{\overline \alpha} \gamma }
t^{-1}_{{\overline \delta} \beta}$ 
($i\equiv \alpha{\overline \beta} \; , \;  j\equiv\gamma{\overline \delta}$)
and  
${\tilde K}_{{\overline \alpha} \beta} = t^{-1}_{{\overline \alpha} \beta}$
respectively, where 
$t \equiv t^{\alpha {\overline \beta}} \equiv
(T+T^{\dagger})^{\alpha {\overline \beta}}$ 
is a hermitian matrix. In our general
notation this implies that the matrix $P$ is given by    
$P^{i\bar j}=(t^{1/2})^{\gamma\bar \alpha}(t^{1/2})^{\beta\bar \delta}$
and the SUSY-breaking parametrization becomes    
$F^S = \sqrt{3} m_{3/2} \sin\theta (S+S^*)e^{-i \gamma_S}$, 
$F = \sqrt{3} m_{3/2} \cos\theta t^{1/2}\Theta t^{1/2}$, where
$\Theta$ is a matrix satisfying 
${\rm  {Tr}} \Theta \Theta^{\dagger} = 1$.}. $P$ is a matrix canonically
normalizing the moduli 
fields\footnote{$P$ can be written as $P=U{\hat K}_{d}^{-1/2}$,
where 
$U$ is a unitary matrix which diagonalizes 
$\hat{K}\equiv K_{{\bar i }j}$, 
$U^\dagger\hat{K}U={\hat K}_{d}$.},
i.e. $P^\dagger\hat{K}P=1$ where $1$ stands for the unit matrix, 
the angle $\theta $ and the complex parameters 
$\Theta _{\bar j}$ just parametrize the
direction of the Goldstino in the $S,T_i$ field space
and $\sum _j \Theta_j^*\Theta_{\bar j}=1$.
We have also allowed for the possibility of
some complex phases which could be relevant
for the CP structure of the theory. This parametrization has the virtue that
when we plug it in the general form of the SUGRA scalar potential
eq.(\ref{pot}), the cosmological constant
\begin{equation}
V_0 = (S+S^*)^{-2}|F^S|^2 + {F}^{\bar i}\hat{K}_{\bar ij}F^j - 3m_{3/2}^2\ ,
\end{equation}
vanishes by
construction. Notice that such a phenomenological approach allows us
to `reabsorb' (or circumvent) our ignorance about the (nonperturbative)
$S$- and $T_i$- dependent part of the superpotential, which is
responsible for SUSY breaking \cite{Brignole}.
Plugging
eq.(\ref{parametrization}) into 
eqs.(\ref{mmatrix},\ref{mmatrix2},\ref{gauginos})
one finds the following results 
\begin{eqnarray}
m'^2_{\bar\alpha\beta} &=& m_{3/2}^2\left[ \tilde{K}_{\bar\alpha\beta}
+3\cos^2\theta\Theta^*_k{P^\dagger}^{k\bar i}
 \left(\partial_{\bar i}\tilde{K}_{\bar\alpha\gamma}\tilde{K}^{\gamma\bar\delta}
 \partial_j\tilde{K}_{\bar\delta\beta}
-\partial_{\bar i}\partial_j\tilde{K}_{\bar\alpha\beta}\right)
P^{j\bar l}\Theta_{\bar l} \right]\ ,
\label{scalar.mass.square}
\\
A'_{\alpha\beta\gamma} &=& -\sqrt3m_{3/2}e^{K/2}\left[
\sin\theta e^{-i\gamma_S}Y_{\alpha\beta\gamma} \ 
+\cos\theta P^{i\bar j}\Theta_{\bar j}
\right. \nonumber\\ && \left. \times
\left\{\left(
{\tilde K^{{ \delta} {\overline{\rho}} }}
\partial_i {\tilde K_{{\overline{\rho}}{ \alpha}}} Y_{\delta\beta\gamma}
+(\alpha \leftrightarrow \beta)+(\alpha \leftrightarrow \gamma)\right)
-\hat{K}_iY_{\alpha\beta\gamma}-\partial_i Y_{\alpha\beta\gamma}\right\}
\right]\ ,
\label{trilinear}
\\
M_a &=& \sqrt3m_{3/2}\sin\theta e^{-i\gamma_S}\ ,
\label{gaugino.mass}
\end{eqnarray}
where the tree-level gaugino masses are independent of the moduli sector
due to the fact that the tree-level gauge kinetic function is given
for any 4-D String by  
$f_a=k_aS$ with $k_a$ the Kac-Moody level of the gauge factor.

As we mentioned above, the parametrization of the auxiliary field VEVs
was chosen in such a way to guarantee the automatic vanishing of
the VEV of the scalar potential ($V_0=0$). If the value of $V_0$
is not assumed to be zero
the above formulae 
(\ref{scalar.mass.square},\ref{trilinear},\ref{gaugino.mass}) 
are modified in the 
following simple way.
One just has to replace $m_{3/2}\rightarrow Cm_{3/2}$,
where $|C|^2=1+V_0/3m_{3/2}^2$. In addition, the formula for 
$m'^2_{\bar\alpha\beta}$
gets an additional contribution given by 
$2m_{3/2}^2(|C|^2-1)\tilde{K}_{\bar\alpha\beta}
=(2V_0/3)\tilde{K}_{\bar\alpha\beta}$.

The soft term formulae above 
(\ref{scalar.mass.square},\ref{trilinear},\ref{gaugino.mass}) 
are valid for any compactification scheme. In addition one is tacitally
assuming that the tree-level K\"ahler potential and $f_a$-functions
constitute a good approximation. In fact, the effects of the one-loop
corrections will in general be negligible except for those corners
of the Goldstino directions in which the tree-level soft terms
vanish. As discussed in ref.\cite{Brignole2} this situation would be
a sort of fine-tuning.

In order to obtain more concrete expressions for the bosonic soft terms
(we recall that gaugino masses (\ref{gaugino.mass})
are compactification-scheme dent) one needs some information
about the K\"ahler potential $K$.
In the following we will concentrate on 
CY compactifications 
where this type of information is known.

\section{Calabi-Yau compactifications}

\subsection{General characteristics}

Let us summarize some generic features of effective SUGRAs coming
from (2,2) CY compactifications\footnote{For a review see
\cite{Kpotentials} and references therein.}. 
The gauge group is $E_6 \times E_8$ and 
the matter fields in ${\bf 27}$ representations of $E_6$ are in one-to-one
correspondence with the $T_i$ moduli (i=1,..,m), whereas the 
$\overline{\bf 27}$ 
representations of $E_6$ are in one-to-one correspondence with the $U_k$
moduli (k=1,..,n). The tree-level K\"ahler potential is 
\begin{eqnarray}
K &=& -\log(S+S^*) + \hat{K}_1(T_i,T_i^*) + \hat{K}_2(U_k,U_k^*)
\nonumber\\ &&
+ \tilde{K}_{1\bar ij}{{\bf 27}^*}^{\bar i}{\bf 27}^j
+ \tilde{K}_{2\bar kl}{\overline{\bf 27}^*}^{\bar k}\overline{\bf 27}^l
+\ (Z_{{i}{k}}{\bf 27}^i{\overline{\bf 27}}^k\ +\ h.c. \ )\ ,
\label{kahlcy}
\end{eqnarray}
where we have included the last term for completeness since it has
recently been realized that it do appears in some CY 
compactifications \cite{Kaplunovsky, Ferrara}. Since its relation with
the phenomenologically required $B$ term of the MSSM is very model 
dependent we will not study it in what follows.
The matter metrics are related to moduli K{\"a}hler potentials 
as \cite{Kpotentials}
\begin{equation}
\tilde{K}_{1\bar ij} = 
\frac{\partial^2\hat{K}_1}{\partial T^*_{i}\partial T_j}
\exp\frac13(\hat{K}_2-\hat{K}_1)
\hbox{\ \ ;\ \ }
\tilde{K}_{2\bar kl} = 
\frac{\partial^2\hat{K}_2}{\partial U^*_{k}\partial U_l}
\exp\frac13(\hat{K}_1-\hat{K}_2)\ .
\label{rollo}
\end{equation}
On the other hand, $\hat{K}_{1}$ and $\hat{K}_{2}$ 
are completely determined in terms of two holomorphic functions, the
prepotentials 
${\cal F}_1(T)$ and ${\cal F}_2(U)$, one for each type of moduli, 
as \cite{varios,Kpotentials}
\begin{eqnarray}
\hat{K}_{1,2} &=& -\log\hat{K}'_{1,2}\ , 
\label{kmatter}
\\
\hat{K}'_1 &=&
(t_l-t^*_l)(\partial_l{\cal F}_1+\partial_{\bar l}{\cal F}_1^*)
-2({\cal F}_1-{\cal F}_1^*) \ ,
\label{kmatter55}
\\
\hat{K}'_2 &=&
(u_l-u^*_l)(\partial_l{\cal F}_2+\partial_{\bar l}{\cal F}_2^*)
-2({\cal F}_2-{\cal F}_2^*) \ .
\label{kmatter56}
\end{eqnarray}
where the variables $t=iT$, $u=iU$ are used. 
The same property follows for 
${\bf 27}^3$ and $\overline{\bf 27}^3$ Yukawa couplings
\begin{eqnarray}
Y_{1ijk} &=& \partial_i\partial_j\partial_k{\cal F}_1\ ,
\label{Yukawas}
\\
Y_{2ijk} &=& \partial_i\partial_j\partial_k{\cal F}_2\ .
\label{Yukawas2}
\end{eqnarray}
Finally, using eqs.(\ref{kmatter},\ref{rollo}), 
the moduli and matter metrics associated with ${\bf 27}$ representations 
can be written as
\begin{eqnarray}
\hat{K}_{1\bar ij} &=& \hat{K}'^{-2}_1X_{1\bar ij}\ , 
\label{nada}\\
\tilde{K}_{1\bar ij} &=& \hat{K}'^{-5/3}_1\hat{K}'^{-1/3}_2X_{1\bar ij}\ ,
\label{algo}
\end{eqnarray}
where
\begin{equation}
X_{1\bar ij} = \partial_{\bar i}\hat{K}'_1\partial_j\hat{K}'_1
        -\hat{K}'_1\partial_{\bar i}\partial_j\hat{K}'_1\ .
\label{X_metric}
\end{equation}

The metrics of $\overline{\bf 27}$ representations
have exactly the same form with the obvious replacement $1\leftrightarrow 2$.

\subsection{Soft terms}

Let us apply the previous generic CY 
formulae (\ref{nada},\ref{algo},\ref{X_metric})
and the parametrization introduced above 
(\ref{scalar.mass.square},\ref{trilinear}) to the computation of the
soft terms. We will concentrate first on those which are associated with
${\bf 27}$ representations. The results are
\begin{eqnarray}
m'^2_{\bar kl} &=& m_{3/2}^2\left[ \tilde{K}_{1\bar kl}
+\cos^2\theta\left\{ 
-\left(1+4\sum_i \Theta_{1i}^*\Theta_{1\bar i}\right)
\tilde{K}_{1\bar kl} \ \right. \right.
\nonumber\\ &&
\left. \left.
     +3\hat{K}'^{-5/3}_1\hat{K}'^{-1/3}_2 
\Theta^*_{1m}{P_1^\dagger}^{m\bar i}\left(
\partial_{\bar i}X_{1\bar kn} X_1^{n\bar o}\partial_jX_{1\bar ol}
- \partial_{\bar i}\partial_j X_{1\bar kl}
\right) P_1^{j\bar r}\Theta_{1\bar r}\right\}\right]\ ,
\label{masaprima} 
\\
A'_{jkl} &=& -\sqrt3m_{3/2}e^{K/2}\left[
\sin\theta e^{-i\gamma_S}Y_{1jkl}
+\cos\theta P_1^{i\bar m}\Theta_{1\bar m}
\right. \nonumber\\ && \left. \times
\left\{\left(X_1^{n\bar o}
\partial_i X_{1\bar oj}Y_{1nkl}
+(j \leftrightarrow k)+(j \leftrightarrow l)\right)
-4\hat{K}'^{-1}_1\hat{K}'_{1i}Y_{1jkl}-\partial_i Y_{1jkl}
\right\}\right] 
\label{trilinearprima}
\end{eqnarray}
where we use latin indices for both matter and moduli fields reflecting
the one-to-one correspondence between them.
Notice that, since the two types of moduli form separate moduli spaces
and the K\"ahler function is
$\hat{K}=\hat{K}_1(T_i,T_i^*) + \hat{K}_2(U_k,U_k^*)$
we have taken the matrices $P$ and $\Theta$ of eq.(\ref{parametrization}) as
\begin{equation}
P=\left( \begin{array}{cc} P_1^{i\bar j} & 0 \\
                           0             & P_2^{k\bar l} \end{array}
  \right)
\hbox{\ \ ;\ \ }
\Theta=\left( \begin{array}{c} \Theta_{1\bar j} \\
                               \Theta_{2\bar l} \end{array}
  \right)
\end{equation}
where $P_{1,2}$ are the matrices canonically normalizing the moduli fields,
i.e. $P_{1,2}^\dagger\hat{K}_{1,2}P_{1,2}=1$, and 
$\sum _j \Theta_{1j}^*\Theta_{1\bar j}+\sum _l \Theta_{2l}^*\Theta_{2\bar l}=1$.


Before canonically normalizing (through a ``rotation'') the matter fields we
already see that the expression 
for CY trilinear soft terms is in general very complicated. 
On the one hand, as we will see below,
the instanton contribution to $Y_{1jkl}$ 
implies that $\partial_iY_{1jkl}\ne 0$. On the other hand,
all Yukawa couplings between ${\bf 27}'s$ are allowed\footnote{For example,
in CY compactifications with two moduli, the four possible different Yukawa
couplings $Y_{jkl}$ with $j,k,l=1,2$
are in general non-vanishing \cite{Katz,Lectures}.}. 
As a consequence, it is not possible in 
general to 
factorize out the Yukawa coupling $Y_{1jkl}$ in (\ref{trilinearprima})
since the terms contained in parenthesis are not proportional to it.
In addition, universality of trilinear parameters is lost.

Let us focus now on soft scalar masses eq.(\ref{masaprima}). If the matter
fields are canonically normalized with the matrix $Q_1$, 
$Q^{\dagger}_1\tilde {K}_1Q_1=1$ 
where $\tilde K_1\equiv {\tilde K}_{1\bar ij}$,
and the matrix $R_1$ is defined such that $R^{\dagger}_1 {X}_1R_1=1$ where 
$X_1\equiv X_{1\bar ij}$ (see eq.(\ref{X_metric})), then the relations
$Q_1=\hat K'^{5/6}_1\hat K'^{1/6}_2R_1$, $P_1=\hat K'_1R_1$ are fulfilled.
Now, taking into account these relations, the normalized soft mass
matrix can be written as 
\begin{equation}
m_{q\bar p}^2 = m_{3/2}^2
\left[\delta_{q\bar p}+\cos^2\theta\left(-\delta_{q\bar p}+
\Delta_{q\bar p}(T_i, T^*_i, \Theta_{\bar i}, \Theta^*_i)\right)
\right]\ , 
\label{pelma2}
\end{equation}
where
\begin{equation}
\Delta_{q\bar p} = -4\delta_{q\bar p}\sum_i \Theta_{1i}^*\Theta_{1\bar i}
     +3\hat{K}'^2_1 {R^\dagger}_1^{q\bar k}\Theta^*_{1m}{R^\dagger}_1
^{m\bar i}\left(
            \partial_{\bar i}X_{1\bar kn} X_1^{n\bar o}
               \partial_jX_{1\bar ol}-\partial_{\bar i}\partial_j 
X_{1\bar kl}
     \right) R_1^{j\bar r}\Theta_{1\bar r}R_1^{l\bar p}\ .
\label{pelma}
\end{equation}
As in the case of trilinear parameters, the expression for soft masses
is very complicated. An interesting question related to flavour changing
issues concerns the degree of degeneracy among the eigenvalues of the matrix
$m_{q\bar p}^2$. In general, $\Delta_{q\bar p}$ depends on $\Theta_i$ and
the moduli $T_i$ and therefore will have a generic matrix structure with 
non-degenerate eigenvalues. On the other hand, tachyons may appear.
We will study in more detail these issues in specific multimoduli examples
below.

However, it is worth noticing here that there is an interesting
situation where universality of soft masses and trilinear parameters is 
achieved. Notice that, although the metric $\tilde {K}_1$ (\ref{algo}) 
associated
with {\bf 27}'s has $U$-moduli dependence through the $\hat {K}'^{-1/3}_2$
factor, soft terms in (\ref{masaprima},\ref{trilinearprima})
are independent of $\Theta_{2\bar l}$. The latter are absorbed by the
relation 
$\sum _j \Theta_{1j}^*\Theta_{1\bar j}+\sum _l \Theta_{2l}^*\Theta_{2\bar l}=1$.
The related consequence is that SUSY breaking by dilaton and/or $U$-moduli
sector (but not by the $T$-moduli sector), that is
$\sum _j \Theta_{1j}^*\Theta_{1\bar j}=0$ and 
$\sum _l \Theta_{2l}^*\Theta_{2\bar l}=1$, leads to universal soft-breaking
parameters for {\bf 27}'s
\begin{equation}
\sqrt 3 m=|M|=|A|=\sqrt 3 m_{3/2}\sin \theta\ ,
\end{equation}
while those for $\overline{\bf 27}$'s are complicated.
The soft terms of $\overline{\bf 27}$ representations
have exactly the same form than above 
(\ref{trilinearprima},\ref{pelma2})
with the replacement $1\leftrightarrow 2$.
Obviously, 
the opposite case (SUSY breaking by $T$ moduli) leads to 
universal soft-breaking parameters for $\overline{\bf 27}$'s.

Finally, let us remark that the above results are completely general. They
are valid for any expression of the prepotentials since they were
obtained just using property (\ref{kmatter}).

\subsection{The large-radius limit}

In order to study more concrete features of the soft terms one needs some
information about the prepotentials ${\cal F}_{1,2}$ 
(see eqs.(\ref{kmatter55},\ref{kmatter56})). As explained in the 
introduction, this information is only known in the large-radius limit.
In particular, 
whereas ${\cal F}_2(U)$ is a complicated function
of the complex structure moduli,
${\cal F}_1(T)$ is simply (barring instanton corrections) a
cubic polynomial. 
{}From now on we will concentrate on $T$ moduli associated with ${\bf 27}$
representations. 
We know that one
of them is the ``overall radius $R$'' of the manifold and therefore is
always present. 
A formal large-radius expansion of ${\cal F}_1(T)$ is \cite{HKTY,KT,Font}
\begin{equation}
{\cal F}_1 = k_{ijk}t_it_jt_k + a_{ij}t_it_j + b_it_i + c + {\cal F}_{\rm inst}
\ ,
\label{efes}
\end{equation}
where the imaginary constant $c$ can be identified with a 
$\sigma$-model loop contribution \cite{Parkes} and
the non-perturbative instanton correction 
${\cal F}_{\rm inst}$ is a power series in
$q_i\equiv e^{2\pi it_i}(=e^{-2\pi T_i})$
\begin{equation}
{\cal F}_{\rm inst} = d_iq_i + e_{ij}q_iq_j + \cdots\ .
\label{power}
\end{equation}
Given eqs.(\ref{Yukawas},\ref{efes}), it is clear that only 
${\cal F}_{\rm inst}$ and the real constant $k_{ijk}$ 
enter the {\bf 27}'s Yukawa couplings
\begin{equation}
Y_{1ijk} = 6k_{ijk} + \partial_i\partial_j\partial_k{\cal F}_{\rm inst}\ .
\label{Yukawa_coupling}
\end{equation}
With respect to the moduli K\"ahler potential, plugging
(\ref{efes}) into eq.(\ref{kmatter55}) one can show that the real constants
$a_{ij}$, $b_j$ are irrelevant for it (and therefore irrelevant for
physical quantities). The final result (to first order in instanton 
correction) is
\begin{eqnarray}
\hat{K}'_1 &=& k_{ijk}(T_i+T^*_i)(T_j+T^*_j)(T_k+T^*_k) - 4ic  
\nonumber\\
&& +4d_i[\pi(T_i+T^*_i)+1]e^{-\pi(T_i+T^*_i)} \sin i\pi(T_i-T^*_i)\ .
\label{CYkahlerfunction}
\end{eqnarray}

\subsection*{One-modulus case}

Several examples of CY manifolds which possess only one K\"ahler modulus
can be found in the literature \cite{Candelas}. In these cases the previous
computation of soft terms is simplified since the moduli and matter metrics
are trivially diagonal. 
The study of these examples, although do not lead to any realistic model,
is highly interesting because 
it will allow us to analyze in detail the
modifications produced on soft terms by the $\sigma$-model
contribution and the instanton correction to the K\"ahler potential
(\ref{CYkahlerfunction}). In more realistic models the same type of 
modification will be present\footnote{It is worth noticing here that, as we
will see in the multimoduli case, 
the ``overall-modulus'' limit, where the assumption that all $T_i$
moduli contribute exactly the same to SUSY breaking is made, is 
equivalent to the one-modulus case.}.
Now eqs.(\ref{CYkahlerfunction},\ref{X_metric}) transform in
\begin{eqnarray}
\hat{K}'_1 &=& k(T+T^*)^3 - 4ic
+ 4d\left[\pi(T+T^*)+1\right]e^{-\pi(T+T^*)}\sin i\pi(T-T^*)\ ,
\label{bebo}
\\
X_1 &=& \overline\partial\hat{K}'_1\partial\hat{K}'_1
   -\hat{K}'_1\overline\partial\partial\hat{K}'_1\ ,
\label{vaya}
\end{eqnarray}
and the soft breaking parameters, using 
eqs.(\ref{pelma2},\ref{pelma},\ref{trilinearprima}) 
with $R_1=X^{-1/2}_1$, $P_1=\hat {K}'_1 X^{-1/2}_1$ 
and $\Theta=e^{-i\gamma_T}$, 
are given by
\begin{eqnarray}
m^2 &=& m_{3/2}^2
\left[1+\cos^2\theta\left(-1+\Delta(T, T^*)\right)
\right]\ , \\
A &=& -\sqrt3m_{3/2}
\left[\sin\theta e^{-i\gamma_S}+\cos\theta e^{-i\gamma_T}\omega(T, T^*)
\right]\ ,
\label{bueno}
\end{eqnarray}
where
\begin{eqnarray}
\Delta(T,T^*) &=& -4 + 3\hat{K}'^2_1X^{-2}_1
    (\overline\partial X_1X^{-1}_1\partial X_1-\overline\partial\partial X_1)\ ,
\label{delta_exact}
\\ 
\omega(T,T^*) &=& \hat{K}'_1 X^{-1/2}_1
    (3X^{-1}_1\partial X_1-4\hat{K}'^{-1}_1\partial\hat{K}'_1-Y^{-1}_1
\partial Y_1)\ .
\label{omega_exact}
\end{eqnarray}
Note that we are assuming for simplicity 
vanishing $F$ terms associated with $U$ moduli,
i.e. $\sum_i \Theta_{1i}^*\Theta_{1\bar i}=1$ in eq.(\ref{pelma})
This will be enough to give us an insight into features of soft terms.
In any case, the inclusion of the $F$ terms associated with $U$ moduli
is straightforward.
Since we have only one Yukawa coupling, it is possible 
to factorize it out trivially in (\ref{trilinearprima}).
This has been carried out 
in eq.(\ref{bueno}) where $e^{K/2}$ has also been factorized out.

In ref.\cite{Brignole} the large-radius limit of CY compactifications was
studied in the approximation that only 
the first term in eq.(\ref{bebo}) contributes to the K\"ahler potential. 
The final 
soft terms were
\begin{eqnarray}
m^2 &=& m_{3/2}^2
\left[1-\cos^2\theta\right]\ , 
\label{malo6}
\\
A &=& -\sqrt3m_{3/2}
\left[\sin\theta e^{-i\gamma_S}-\cos\theta e^{-i\gamma_T}
\frac{(T+T^*)}{\sqrt3}\frac{\partial_T Y}{Y}\right]\ .
\label{malo5}
\end{eqnarray}
It is easy to see that 
this result is recovered with our formulae since $\Delta=0$ 
and $\omega=-(\sqrt 3)^{-1}(T+T^*)Y^{-1}\partial_T Y$
when $K=-\log k(T+T^*)^3$.
It can be further argued that, in the large-radius limit, the Yukawa couplings
tend exponentially to constants 
(see eqs.(\ref{power},\ref{Yukawa_coupling})). Then one can take
$\omega \rightarrow 0$ and therefore
\begin{eqnarray}
A &\simeq& -\sqrt3m_{3/2}\sin\theta e^{-i\gamma_S}\ .
\label{malo}
\end{eqnarray}
Using eq.(\ref{gaugino.mass}) 
the following relations between soft terms are fulfilled
\begin{equation}
\sqrt 3 m=|M|\simeq |A|\ .
\label{relations}
\end{equation}

Let us now study the departure from the previous results 
(\ref{malo6},\ref{malo5})
when the
most general case, including the terms proportional to $c$ and $d$
in (\ref{bebo}), is considered. Due to these $\sigma$-model
and instanton contributions the analytic calculation is more involved.
Using again eqs.(\ref{delta_exact},\ref{omega_exact}) 
and expanding the quantities in terms of $1/(T+T^*)$
we obtain at the leading order\footnote{We note for completeness that the
matter metric (\ref{algo}) is 
$\tilde {K}_1 = k^{1/3}\hat {K}'^{-1/3}_2 
[{3}{(T+T^*)^{-1}} - {11C}{(T+T^*)^{-4}}
+{4\pi D}{[\pi(T+T^*)]^{-2}} e^{-\pi(T+T^*)}\sin i\pi(T-T^*)]$.}
\begin{eqnarray}
\Delta(T,T^*) &=& \frac{40C}{(T+T^*)^3}
    +\frac{8D}{3\pi}e^{-\pi(T+T^*)}\sin i\pi(T-T^*)\ ,
\label{delta_approximate} \\
\omega(T,T^*) &=& \frac{10\sqrt3C}{(T+T^*)^3}
    +\frac{4D}{\sqrt3}e^{-2\pi T^*}
    -\frac{(T+T^*)}{\sqrt3}\frac{\partial_T Y}{Y}\ ,
\label{omega_approximate}
\end{eqnarray}
where $C\equiv -4ic/k$ and $D\equiv 4\pi^3d/k$.
As discussed above, the third term in (\ref{omega_approximate}) is
negligible.
These are the moduli-dependent modifications to soft terms 
arising from $\sigma$-model and instanton contributions.
They are also model dependent due to the factors $k$, $c$ and $d$. 
For the four one-modulus models classified in ref.\cite{Candelas}
these factors have been computed. In particular \cite{KT,Font}
\begin{equation}
(k,c) = (5/6,{25i}\zeta(3)/{\pi^3})
\hbox{\ ;\ }
(1/2,{51i}\zeta(3)/{2\pi^3})
\hbox{\ ;\ }
(1/3,{37i}\zeta(3)/{\pi^3})
\hbox{\ ;\ }
(1/6,{36i}\zeta(3)/{\pi^3})
\label{numbers}
\end{equation}
and $d$ is of order $100$ \cite{HKTY}. 
Using these numbers we can see that the orders 
of magnitude of $C$ and $D$ are
$10$ and $10^4$ respectively, and therefore, depending on the value of 
${\rm Re\:}T$, the corrections 
(\ref{delta_approximate},\ref{omega_approximate}) might be 
sizeable\footnote{Notice that in the modulus-dominated SUSY-breaking 
scenario, $\sin\theta \rightarrow 0$, these 
corrections may be very important in general since the tree-level
soft terms are vanishing. This was first pointed out in the context of
SUSY breaking by gaugino condensation in ref.\cite{Kiwoon}.}. 
For instance, taking ${\rm Re\:}T=5$, we obtain 
(using the average values $C=10$ and $D=10^4$)
that the first term in
eqs.(\ref{delta_approximate}) and (\ref{omega_approximate}) acquires 
the values
$0.4$ and $0.17$ respectively whereas the second one acquires the values
$-2\times 10^{-10}\sin 2\pi {\rm Im} T$ and 
$5\times 10^{-10} e^{i2\pi {\rm Im} T}$ 
respectively. Clearly the latter, 
which is due to instanton corrections, is negligible 
as was to be expected since it is 
exponentially suppressed in the large-radius limit. 
However, the $\sigma$-model contribution
in this ``average'' model is large and produces the following soft terms
\begin{eqnarray}
m^2 &\simeq& m_{3/2}^2
\left[1-0.6\cos^2\theta\right]\ , 
\label{malo3}
\\
A &\simeq& -\sqrt3m_{3/2}
\left[\sin\theta e^{-i\gamma_S}+0.17\cos\theta e^{-i\gamma_T}
\right]\ ,
\label{malo4}
\end{eqnarray}
to be compared with those of eqs.(\ref{malo6},\ref{malo}) in the
$\Delta=\omega=0$ limit. Unlike what happens in 
that limit (see eq.(\ref{relations})) now there is, in principle, 
the possibility of having scalars heavier than gauginos. In this 
``average'' model the necessary condition is $\cos^2\theta > 0.83$, 
and in general it will be 
\begin{equation}
\cos^2\theta > \frac{2}{2+\Delta}\ .
\label{pesados}
\end{equation}

In Figure~1 we show 
the behavior of $\Delta$ and $|\omega|$ as functions of
${\rm Re\:}T$ for the last model of eq.(\ref{numbers}).
The solid lines correspond to the exact results
eqs.(\ref{delta_exact},\ref{omega_exact}) whereas the dashed lines
correspond to their approximations
eqs.(\ref{delta_approximate},\ref{omega_approximate}). These
approximations are quite good
and for large enough values of ${\rm Re\:}T$ both lines coincide.
Let us remark that $\omega$, although we neglect the Yukawa coupling 
contribution, is a complex quantity due to the instanton correction. 
In the figure we take the particular value ${\rm Im} T=1/4$ for which
$\Delta$ in (\ref{delta_approximate}) is maximum, but different values will
not modify the figure due to the smallness of the instanton correction.
In this CY model the departure from the 
$\Delta=\omega=0$ limit is larger than in the ``average'' model studied above.
In particular, for 
${\rm Re\:}T=5$ as above we get $\Delta=1.62$ and $|\omega|=0.64$.
For instance, $m^2\simeq m_{3/2}^2[1+0.62\cos^2\theta]$ 
and scalars will be heavier than gauginos if $\cos^2\theta>0.55$.
Finally, the dotted lines represent the absolute value of 
instanton contributions inside  
eqs.(\ref{delta_exact},\ref{omega_exact}). As mentioned above 
they are essentially negligible.

\subsection*{Multimoduli case}

As we already mentioned above, in the case of
CY compactifications with several moduli, the metrics are not diagonal
and therefore a general analysis becomes very involved. We
will try to study what new features can appear in this situation. 
Since we are mainly interested in the consequences of off-diagonal metrics,
in order to simplify the analysis, we will consider in
(\ref{CYkahlerfunction}) only those terms which are 
relevant for this issue. Thus we
are left with the cubic terms.
The inclusion of $\sigma$-model and instanton contributions is straightforward
following the lines of the previous section.

Let us first consider the two illustrative two-moduli CY 
manifolds of ref.\cite{HKTY} whose prepotentials are
\begin{eqnarray}
{\cal F} &=& \frac{5}{6}(t_1^3+t_2^3) + 5(t_1^2t_2+t_1t_2^2)\ , \\
{\cal F'} &=& \frac{1}{3}t_1^3 + 4t_1^2t_2 + 3t_1t_2^2\ .
\end{eqnarray}
Taking into account the general formulae of section 3.1 and 
eq.(\ref{CYkahlerfunction}), 
this information
is enough
in order to discuss the soft scalar masses (\ref{pelma2}).
The matrix $\Delta_{i\bar j}$ (\ref{pelma}) with $i, j=1, 2$ is extremely
involved and depends on moduli $T_i$ as well
as $\Theta_i$. In fact, 
the moduli dependence of soft breaking parameters appears through the ratio 
$T_2/T_1$ due to our parametrization.
In order to get some conclusions we examine $\Delta_{i\bar j}$ 
for specific values of $T_2/T_1$.
In the case of ${\cal F}$ with $T_2/T_1=1$ and $T_2/T_1=2$, 
$\Delta_{i\bar j}$ is simplified to
\begin{equation}
\left(\begin{array}{cc}
 0 & -4\Theta_{\bar 1}\Theta_{\bar 2} \\
 -4\Theta_{\bar 1}\Theta_{\bar 2} & -3\Theta_{\bar 2}^2
\end{array}\right)
\label{primera}
\end{equation}
and
\begin{equation}
\left(\begin{array}{cc}
-0.7281\Theta_{\bar 1}^2 +2.262\Theta_{\bar 1}\Theta_{\bar 2} +0.1866\Theta_{\bar 2}^2 &
 1.131\Theta_{\bar 1}^2 -3.627\Theta_{\bar 1}\Theta_{\bar 2}+0.07544\Theta_{\bar 2}^2 \\
 1.131\Theta_{\bar 1}^2 -3.627\Theta_{\bar 1}\Theta_{\bar 2}+0.07544\Theta_{\bar 2}^2 &
 0.1866\Theta_{\bar 1}^2+0.1509\Theta_{\bar 1}\Theta_{\bar 2}  -2.375\Theta_{\bar 2}^2
\end{array}\right),
\label{segunda}
\end{equation}
respectively, where the parameters $\Theta_{\bar j}$ and the 
$F$ terms associated with $U$ moduli have been taken 
real and vanishing respectively for simplicity. 
${\cal F}$ possesses a symmetry under the exchange of $t_1$ 
and $t_2$.  For example, $T_2/T_1=2$ and $T_2/T_1=1/2$ yield the same result.  
In the case of ${\cal F'}$, e.g. with $T_2/T_1=1$ we obtain 
\begin{equation}
\left(\begin{array}{cc}
 -0.1712\Theta_{\bar 1}^2+1.127\Theta_{\bar 1}\Theta_{\bar 2} +0.1267\Theta_{\bar 2}^2 &
  0.5637\Theta_{\bar 1}^2-3.747\Theta_{\bar 1}\Theta_{\bar 2}-0.1945\Theta_{\bar 2}^2 \\
  0.5637\Theta_{\bar 1}^2-3.747\Theta_{\bar 1}\Theta_{\bar 2}-0.1945\Theta_{\bar 2}^2 &
  0.1267\Theta_{\bar 1}^2 -0.3891\Theta_{\bar 1}\Theta_{\bar 2} -3.077\Theta_{\bar 2}^2
\end{array}\right).
\label{tercera}
\end{equation}
As mentioned below (\ref{pelma}), $\Delta_{i\bar j}$ has a generic
matrix structure with non-vanishing eigenvalues and therefore
universality is lost in general\footnote{For a discussion of how to
ameliorate the situation concerning FCNC, taking into account the
flavour independent contribution from gauginos to low-energy scalar masses,
see \cite{Brignole,Varioss}.} unless $\cos^2\theta \ll 1$, i.e. the 
dilaton-dominated SUSY-breaking scenario \cite{Kaplunovsky,Brignole}, or
SUSY is broken by $U$-moduli sector. 
E.g., For $\Delta_{i\bar j}$ given by (\ref{primera})
the two squared-mass eigenvalues are
\begin{eqnarray}
m_{1,2}^2 &=& m_{3/2}^2
\left[1+\cos^2\theta\left(-1+\Delta_{1,2}\right)
\right]\ ,
\label{mjj} 
\end{eqnarray}
where
\begin{eqnarray}
\Delta_{1,2}=-\frac{3}{2}\Theta_2^2\pm \sqrt{\frac{9}{4}\Theta_2^4 + 
16 \Theta_1^2 \Theta_2^2}\ .
\label{mjjj}
\label{yoquese}
\end{eqnarray}
For instance, $\Theta_1=\Theta_2=1/\sqrt 2$ implies 
$\Delta_{1,2}=1.39, -2.89$, i.e. 
$m_{1}^2=m_{3/2}^2(1+0.39\cos^2\theta)$ and
$m_{2}^2=m_{3/2}^2(1-3.89\cos^2\theta)$.
We can find the specific $\Theta_i$ at which the eigenvalues become
degenerate (this is of course a sort of fine-tuning). In this case
$\Theta_1=1$, $\Theta_2=0$. For $\Delta_{i\bar j}$ given by 
eq.(\ref{segunda}) (eq.(\ref{tercera})), 
$\Theta_1=0.9541$, $\Theta_2=0.2994$ ($\Theta_1=0.9890$, $\Theta_2=0.1477$).
This requires $\Delta_{i\bar j}=0$, i.e. one is pushed towards the 
``overall-modulus'' limit\footnote{Notice that this limit is 
equivalent to the one-modulus case, see (\ref{malo6}).
We recall that we are not considering in this
computation, for the sake of simplicity, 
the $\sigma$-model and instanton contributions.}, 
$|F^{T_1}|=|F^{T_2}|=\sqrt 3 m_{3/2} \cos\theta /\sqrt 2$ 
(see eq.(\ref{parametrization})).

This is a general result, the universality of soft breaking parameters
in multimoduli cases is achieved by vanishing $\Delta_{i\bar j}$.
The equivalent result for trilinear parameters is that the braces
in (\ref{trilinearprima}) must be vanishing.

Since some of the 
eigenvalues $\Delta_i$ can be positive (see e.g. (\ref{yoquese})),
this open the possibility of having scalars heavier than gauginos depending
on the Goldstino direction. The necessary condition is the same than in
(\ref{pesados}) with $\Delta=\Delta_i$.

Finally, for negative eigenvalues, depending again on the Goldstino
direction, tachyons may appear unless\footnote{It is worth pointing out
here that their possible existence is not necessarily a problem, but
in some cases may be an interesting advantage in order to break extra
gauge symmetries \cite{Brignole2}.}
\begin{equation}
\cos^2\theta < \frac{1}{1-\Delta_i}\ .
\label{tachyons}
\end{equation}

\section{A comparison with orbifold compactifications}

It is interesting to compare the previous results on soft terms with those of 
orbifold compactifications \cite{Brignole,Brignole2}. We recall that 
orbifolds are a special limit of particular CY manifolds.
Let us focus
first on the ``overall-modulus'' case. 
For this class of orbifold models the
K\"ahler potential has the form
\begin{eqnarray}
& K = \
-\log(S+S^*)\ -3\log(T+T^*)\
+\sum_{\alpha}|C_{\alpha}|^2 (T+T^*)^{n_{\alpha}}\ .
\label{kahloverall}
\end{eqnarray}
It is important to remark that, unlike the case of smooth CY models, the
above $T$ dependence does not get corrections from world-sheet
instantons and is equally valid for small and large $T$.
Notice that, with matter fields in the untwisted sector, i.e. modular
weight $n_{\alpha}=-1$, the resulting K\"ahler potential
is analogous to the one obtained in the large-radius limit of CY models
neglecting the $\sigma$-model and instanton corrections 
(see eqs.(\ref{kmatter},\ref{bebo}) and footnote 10). 
As a consequence, in the untwisted sector of orbifolds the soft 
terms are also given by (\ref{malo6},\ref{malo}) with gaugino masses
always bigger than scalar masses (\ref{relations}).
Unlike CY models now eq.(\ref{malo}) is exact since Yukawa couplings
involving untwisted fields are constants and therefore the last term
in (\ref{malo5}) is exactly zero.

As discussed below eq.(\ref{relations}), the situation is qualitatively
different when $\sigma$-model and instanton corrections are included
in CY models. Then, the soft terms are modified due to the non-vanishing
values of $\Delta$ and $\omega$ in 
eqs.(\ref{delta_approximate},\ref{omega_approximate})
allowing e.g. the possibility of scalars heavier than gauginos.

Relaxing the ``overall-modulus'' assumption in orbifolds, some qualitative
changes appear\footnote{For an extensive discussion see \cite{Brignole2}.}.
In the multimoduli case, non-universal soft scalar masses for untwisted
particles are allowed and in fact this will be the most general situation.
Tachyons may also be present. Finally, there may be scalars with mass bigger
than gauginos. However, on average the scalars are lighter than gauginos
since three particles linked via a renormalizable untwisted Yukawa
coupling always fulfill the sum rule
\begin{equation}
m_1^2+m_2^2+m_3^2 = |M|^2\ ,
\label{sumrule}
\end{equation}
as in the ``overall-modulus'' case ($3m^2=|M|^2$). For trilinear parameters
one also has
\begin{equation}
A_{123}=-M\ .
\label{sumrule2}
\end{equation}
In ref.\cite{Brignole2} was shown that these results 
(\ref{sumrule},\ref{sumrule2})
are satisfied even in
the presence of off-diagonal 
metrics. 
Let us recall that although diagonal
metrics is the generic case in most of orbifolds, off-diagonal ones
appear for fields in the untwisted sector of the orbifolds
$Z_3, Z_4, Z'_6$ (see footnote 4).

In CY compactifications where off-diagonal metrics are always present, 
some of the
above orbifold properties 
are qualitatively similar. In the previous section we showed that 
non-universality is a natural situation due to the generic 
matrix structure of the soft masses\footnote{However, as showed in section
3.2, in the special situation where only the
$U$ ($T$) moduli contribute to SUSY breaking, the  
${\bf 27}$ ($\overline{\bf 27}$) soft masses and trilinear terms are
universal. This is not true in the case of orbifolds.}. 
Likewise, the presence of tachyons
or scalars heavier than gauginos was allowed. However, the sum rule
(\ref{sumrule}) is violated in general. As an example, we can consider
the two-moduli cases studied in the previous section, where 
$Y_{111}$, $Y_{122}$, $Y_{211}$, $Y_{222}$ are the allowed Yukawa
couplings. 
{}From (\ref{mjj}) we obtain
\begin{eqnarray}
3m_1^2 &=& |M|^2 + 3m_{3/2}^2 \Delta_1\cos^2\theta\ ,
\nonumber 
\\
3m_2^2 &=& |M|^2 + 3m_{3/2}^2 \Delta_2\cos^2\theta\ ,
\nonumber
\\
m_1^2+2m_2^2 &=& |M|^2 + 3m_{3/2}^2 \frac{\Delta_1+2\Delta_2}{3}\cos^2\theta\ ,
\nonumber
\\
m_2^2+2m_1^2 &=& |M|^2 + 3m_{3/2}^2 \frac{\Delta_2+2\Delta_1}{3}\cos^2\theta\ ,
\end{eqnarray}
where, taking into account (\ref{mjjj}), is easy to see that 
$3m_1^2>|M|^2$; $3m_2^2, m_1^2+2m_2^2 < |M|^2$ and both possibilities are
allowed for $m_2^2+2m_1^2$
depending on the $\Theta_{1,2}$ values. Due to these two possibilities,
we cannot say as in the case of orbifolds that on average the scalars
are lighter than gauginos. This is an interesting novel fact.

Likewise (\ref{sumrule2}), which is fulfilled in orbifolds due to the
antisymmetric property of Yukawa couplings, no longer is true 
in CY compactifications as discussed
below eq.(\ref{trilinearprima}).

\section{Conclusions}

The study of the soft SUSY-breaking terms coming from string compactifications
is highly important since they determine the low-energy phenomenology of these
theories. We have carried out this study for some specially interesting
and complicated spaces. In particular, Calabi-Yau manifolds have
been analyzed with all the detail allowed by the present state-of-the-art
technology. As discussed in the introduction only the large-radius limit
of (2,2) theories is really available. 

After providing a parametrization
of the soft terms in terms of ``Goldstino angles'' for the most general
case of off-diagonal metrics, we have computed them for Calabi-Yau spaces.
Although these results are general in the sense that they do not depend
on the size of the manifold, in order to obtain more concrete features
we have also worked in the large-radius limit where the prepotential is
known.
As mentioned above, the soft terms 
show formidable complexity mainly due to the
$\sigma$-model and instanton contributions to the K\"ahler potential
as well as the existence of off-diagonal moduli and matter metrics.
Although in the large-radius limit instanton contributions are negligible
since they are exponentially suppressed, $\sigma$-model corrections to the
soft terms may be important depending on the value of ${\rm Re} T$.
On the other hand, very specific features appear due to the presence
of off-diagonal metrics. For example, soft scalar masses have a generic
matrix structure giving rise to non-universality in general.
However, there is an interesting situation, SUSY breaking 
by $U$ ($T$)-moduli sector,
where universality for ${\bf 27}$ ($\overline{\bf 27}$) representations 
is achieved. 
Tachyons may also appear.
Other 
features are related with the possibility of having 
scalars lighter or heavier than gauginos depending on the Goldstino direction.
In this sense, sum rules found in orbifold models, which imply that on 
average the scalars are lighter than gauginos, can be violated in 
Calabi-Yau manifolds.

\section*{Acknowledgements}

We thank L.E. Ib\'a\~nez for useful comments.



\def\MPL #1 #2 #3 {{\em Mod.~Phys.~Lett.}~{\bf#1}\ (#2) #3 }
\def\NPB #1 #2 #3 {{\em Nucl.~Phys.}~{\bf B#1}\ (#2) #3 }
\def\PLB #1 #2 #3 {{\em Phys.~Lett.}~{\bf B#1}\ (#2) #3 }
\def\PR  #1 #2 #3 {{\em Phys.~Rep.}~{\bf#1}\ (#2) #3 }
\def\PRD #1 #2 #3 {{\em Phys.~Rev.}~{\bf D#1}\ (#2) #3 }
\def\PRL #1 #2 #3 {{\em Phys.~Rev.~Lett.}~{\bf#1}\ (#2) #3 }
\def\PTP #1 #2 #3 {{\em Prog.~Theor.~Phys.}~{\bf#1}\ (#2) #3 }
\def\RMP #1 #2 #3 {{\em Rev.~Mod.~Phys.}~{\bf#1}\ (#2) #3 }
\def\ZPC #1 #2 #3 {{\em Z.~Phys.}~{\bf C#1}\ (#2) #3 }

\newpage

\section*{Figure Captions}

\begin{description}
\item[Figure 1] $\Delta(T,T^*)$ and $|\omega(T,T^*)|$ as 
functions of ${\rm Re\:}T$.
The solid lines correspond to the exact results. The dashed lines correspond
to the approximate results. Finally, the dotted lines represent the
absolute value of instanton corrections.
 
\end{description}

\vskip 12mm

\begin{center}
\setlength{\unitlength}{0.9mm}
\begin{picture}(130,200)(0,0)
\put(0,115){\epsfig{file=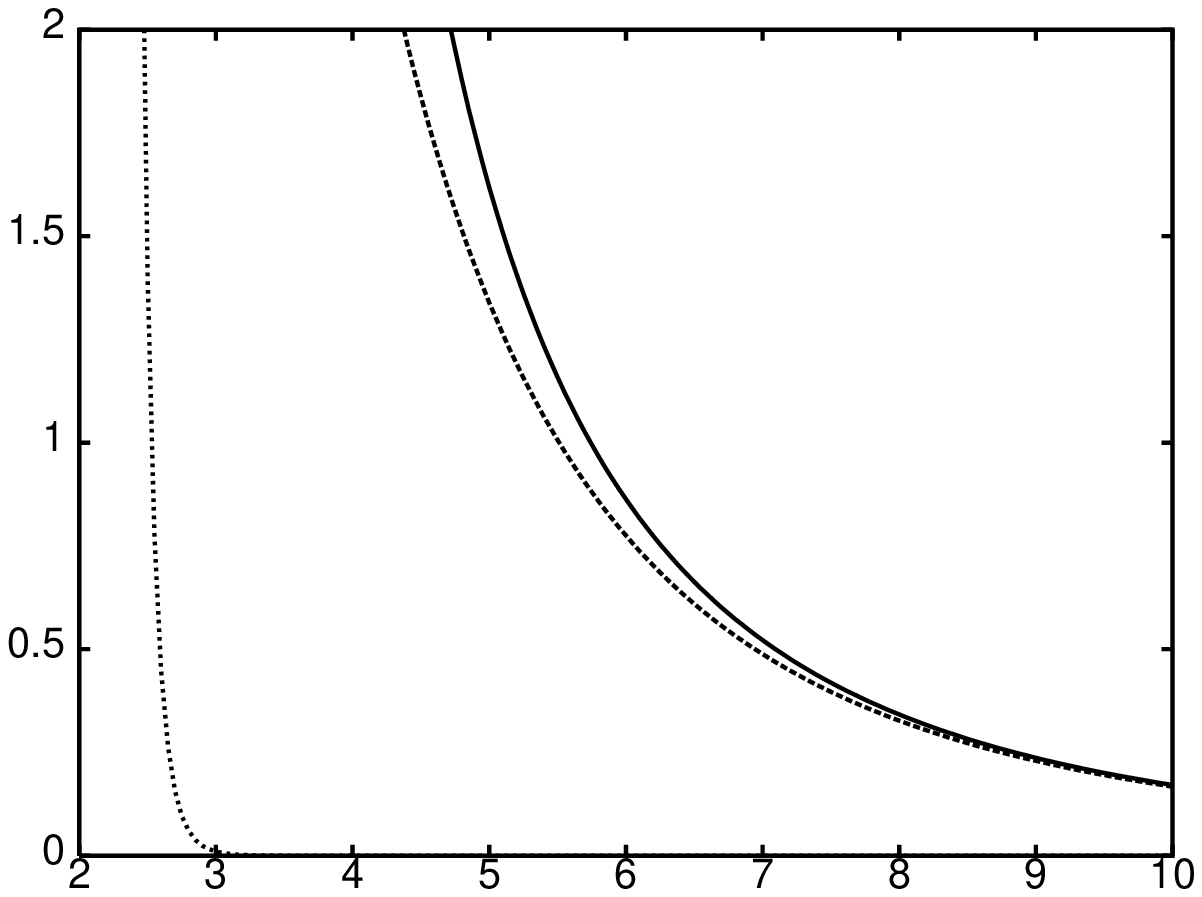,width=120mm}}
\put(63,110){\Large${\rm Re}\:T$}
\put(0,160){\Large$\Delta$}
\put(0,10){\epsfig{file=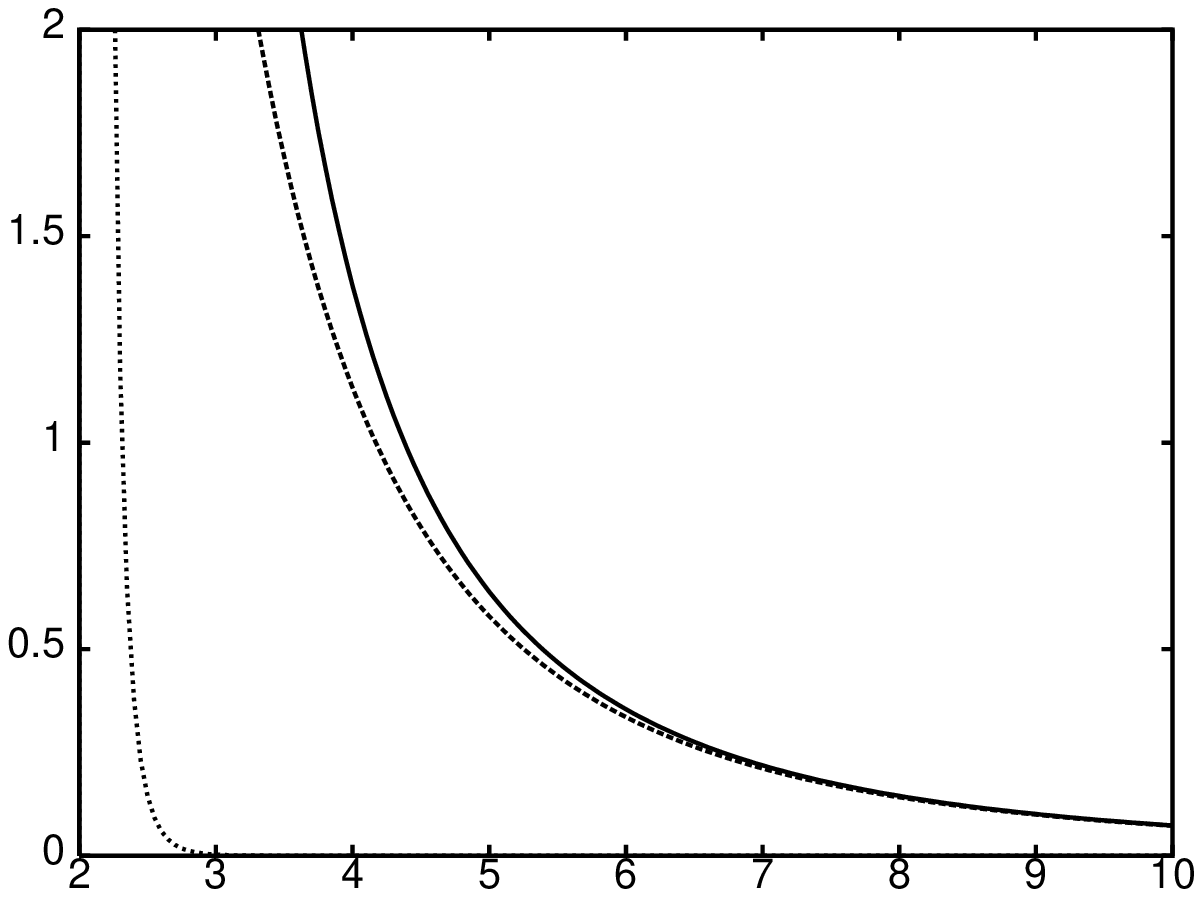,width=120mm}}
\put(63,5){\Large${\rm Re}\:T$}
\put(0,55){\Large$|\omega|$}
\end{picture}
\end{center}

\end{document}